\documentclass[aps,prl,twocolumn,nofootinbib,floatfix,superscriptaddress,showpacs,amssymb,amsmath]{revtex4}
\usepackage{graphicx}
\begin{document}

\title{Persistence of odd-even staggering in charged fragment 
yields \\ from the $^{112}$Sn+$^{58}$Ni collision at 35 MeV/nucleon}

\author{G.~Casini}
\affiliation{Sezione INFN di Firenze, Via G. Sansone 1, I-50019 Sesto Fiorentino, Italy}

\author{S.~Piantelli}
\affiliation{Sezione INFN di Firenze, Via G. Sansone 1, I-50019 Sesto Fiorentino, Italy}

\author{P.R.~Maurenzig}
\affiliation{Sezione INFN di Firenze, Via G. Sansone 1, I-50019 Sesto Fiorentino, Italy}
\affiliation{Dipartimento di Fisica, Univ. di Firenze, Via G. Sansone 1, I-50019 Sesto Fiorentino, Italy}

\author{A.~Olmi}
\thanks{corresponding author} \email[e-mail:]{olmi@fi.infn.it}
\affiliation{Sezione INFN di Firenze, Via G. Sansone 1, I-50019 Sesto Fiorentino, Italy}

\author{L.~Bardelli}
\affiliation{Sezione INFN di Firenze, Via G. Sansone 1, I-50019 Sesto Fiorentino, Italy}
\affiliation{Dipartimento di Fisica, Univ. di Firenze, Via G. Sansone 1, I-50019 Sesto Fiorentino, Italy}

\author{S.~Barlini}
\affiliation{Sezione INFN di Firenze, Via G. Sansone 1, I-50019 Sesto Fiorentino, Italy}
\affiliation{Dipartimento di Fisica, Univ. di Firenze, Via G. Sansone 1, I-50019 Sesto Fiorentino, Italy}

\author{M.~Benelli}
\affiliation{Sezione INFN di Firenze, Via G. Sansone 1, I-50019 Sesto Fiorentino, Italy}
\affiliation{Dipartimento di Fisica, Univ. di Firenze, Via G. Sansone 1, I-50019 Sesto Fiorentino, Italy}

\author{M.~Bini}
\affiliation{Sezione INFN di Firenze, Via G. Sansone 1, I-50019 Sesto Fiorentino, Italy}
\affiliation{Dipartimento di Fisica, Univ. di Firenze, Via G. Sansone 1, I-50019 Sesto Fiorentino, Italy}

\author{M.~Calviani}
\thanks{present address:
       CERN, Geneva, Switzerland}
\affiliation{Sezione INFN di Firenze, Via G. Sansone 1, I-50019 Sesto Fiorentino, Italy}
\affiliation{Dipartimento di Fisica, Univ. di Firenze, Via G. Sansone 1, I-50019 Sesto Fiorentino, Italy}

\author{P.~Marini}
\thanks{present address:
        GANIL, CEA/DSM-CNRS/IN2P3, Caen, France}
\affiliation{Sezione INFN di Firenze, Via G. Sansone 1, I-50019 Sesto Fiorentino, Italy}
\affiliation{Dipartimento di Fisica, Univ. di Firenze, Via G. Sansone 1, I-50019 Sesto Fiorentino, Italy}

\author{A.~Mangiarotti}
\thanks{present address:
        Laborat\'orio de Instrumenta\c{c}\~{a}o e 
        F\'{\i}sica Experimental de Part\'{\i}culas, 
       Coimbra, Portugal}
\affiliation{Sezione INFN di Firenze, Via G. Sansone 1, I-50019 Sesto Fiorentino, Italy}
\affiliation{Dipartimento di Fisica, Univ. di Firenze, Via G. Sansone 1, I-50019 Sesto Fiorentino, Italy}

\author{G.~Pasquali}
\affiliation{Sezione INFN di Firenze, Via G. Sansone 1, I-50019 Sesto Fiorentino, Italy}
\affiliation{Dipartimento di Fisica, Univ. di Firenze, Via G. Sansone 1, I-50019 Sesto Fiorentino, Italy}

\author{G.~Poggi}
\affiliation{Sezione INFN di Firenze, Via G. Sansone 1, I-50019 Sesto Fiorentino, Italy}
\affiliation{Dipartimento di Fisica, Univ. di Firenze, Via G. Sansone 1, I-50019 Sesto Fiorentino, Italy}

\author{A.A.~Stefanini}
\affiliation{Sezione INFN di Firenze, Via G. Sansone 1, I-50019 Sesto Fiorentino, Italy}
\affiliation{Dipartimento di Fisica, Univ. di Firenze, Via G. Sansone 1, I-50019 Sesto Fiorentino, Italy}

\author{M.~Bruno}
\affiliation{Dipartimento di Fisica dell'Universit\'a and Sezione INFN, 40127 Bologna, Italy}

\author{L.~Morelli}
\affiliation{Dipartimento di Fisica dell'Universit\'a and Sezione INFN, 40127 Bologna, Italy}

\author{V.L.~Kravchuk}
\affiliation{INFN-Laboratori Nazionali di Legnaro, 35020 Legnaro, Italy}

\author{F.~Amorini}     
\affiliation{INFN-Laboratori Nazionali del Sud, Via S. Sofia, Catania, Italy}
\affiliation{Dipartimento di Fisica e Astronomia, Univ. di Catania, Catania, Italy}

\author{L.~Auditore}     
\affiliation{INFN-Gruppo Collegato di Messina and Dipt. di Fisica, Univ. di Messina, Messina, Italy}

\author{G.~Cardella}     
\affiliation{INFN-Sezione di Catania, Via S. Sofia, 95123 Catania, Italy}

\author{E.~De Filippo}   
\affiliation{INFN-Sezione di Catania, Via S. Sofia, 95123 Catania, Italy}

\author{E.~Galichet}     
\thanks{present address:
    Institut de Physique Nucl\'eaire, CNRS/IN2P3, Universit\'e Paris-Sud, Orsay, France}
\affiliation{INFN-Laboratori Nazionali del Sud, Via S. Sofia, Catania, Italy}

\author{E.~La Guidara}   
\affiliation{INFN-Sezione di Catania, Via S. Sofia, 95123 Catania, Italy}

\author{G.~Lanzalone}    
\affiliation{INFN-Laboratori Nazionali del Sud, Via S. Sofia, Catania, Italy}
\affiliation{Facolt\`a di Ingegneria ed Architettura, Universit\`a "Kore" di Enna, Enna, Italy}

\author{G.~Lanzan\'o}      
\thanks{deceased}
\affiliation{INFN-Sezione di Catania, Via S. Sofia, 95123 Catania, Italy}

\author{C.~Maiolino}     
\affiliation{INFN-Laboratori Nazionali del Sud, Via S. Sofia, Catania, Italy}

\author{A.~Pagano}       
\affiliation{INFN-Sezione di Catania, Via S. Sofia, 95123 Catania, Italy}

\author{M.~Papa}         
\affiliation{INFN-Sezione di Catania, Via S. Sofia, 95123 Catania, Italy}

\author{S.~Pirrone}      
\affiliation{INFN-Sezione di Catania, Via S. Sofia, 95123 Catania, Italy}

\author{G.~Politi}       
\affiliation{Dipartimento di Fisica e Astronomia, Univ. di Catania, Catania, Italy}
\affiliation{INFN-Sezione di Catania, Via S. Sofia, 95123 Catania, Italy}

\author{A.~Pop}          
\affiliation{Institute for Physics and Nuclear Engineering, Bucharest, Romania}

\author{F.~Porto}        
\affiliation{INFN-Laboratori Nazionali del Sud, Via S. Sofia, Catania, Italy}
\affiliation{Dipartimento di Fisica e Astronomia, Univ. di Catania, Catania, Italy}

\author{F.~Rizzo}        
\affiliation{INFN-Laboratori Nazionali del Sud, Via S. Sofia, Catania, Italy}
\affiliation{Dipartimento di Fisica e Astronomia, Univ. di Catania, Catania, Italy}

\author{P.~Russotto}      
\affiliation{INFN-Laboratori Nazionali del Sud, Via S. Sofia, Catania, Italy}
\affiliation{Dipartimento di Fisica e Astronomia, Univ. di Catania, Catania, Italy}

\author{D.~Santonocito}  
\affiliation{INFN-Laboratori Nazionali del Sud, Via S. Sofia, Catania, Italy}

\author{A.~Trifir\'o}      
\affiliation{INFN-Gruppo Collegato di Messina and Dipt. di Fisica, Univ. di Messina, Messina, Italy}

\author{M.~Trimarchi}     
\affiliation{INFN-Gruppo Collegato di Messina and Dipt. di Fisica, Univ. di Messina, Messina, Italy}


\date{\today}

\begin{abstract}
Odd-even staggering effects on charge distributions are investigated
for fragments produced in semiperipheral and central collisions 
of $^{112}$Sn+$^{58}$Ni at 35 MeV/nucleon.
For fragments with Z$\leq$16 
one observes a clear overproduction of even charges,
which decreases for heavier fragments.
Staggering persists up to Z $\sim$ 30.
It appears to be substantially independent of the centrality of the collisions,
suggesting that it is mainly related to the last few steps in the decay 
of hot nuclei.
\end{abstract}
\pacs{25.70.-z,25.70.Lm,25.70.Mn,25.70.Pq,29.40.-n}

\maketitle

An enhanced production of even-Z fragments with respect to odd-Z ones 
has been observed since long time in a variety of nuclear reactions
(\cite{Yang,Winchester,Ricciardi,geraci,napolitani,dago} 
and references therein).
The enhancement in the yield of even elements over the 
neighboring odd ones is usually of the order of few tens per cent 
for the lightest elements and it rapidly decreases with increasing Z.
Recent experiments \cite{Ricciardi,dago,lombardo}, 
performed with detectors providing good isotopic resolution,
revealed a rather complex behavior.
It was found that even-mass fragments display indeed an odd-even staggering 
with an enhanced yield of even-Z nuclei.
This effect is particularly prominent for N=Z fragments, where it becomes 
of the order of 50\%.
On the contrary, odd-mass fragments display a weaker reverse staggering,
which favors odd-Z nuclei \cite{Ricciardi,dago}.
The overall odd-even effect usually reported in literature results from 
the superposition of these different behaviors.
By comparing colliding systems with different N/Z, it was concluded 
that the staggering on Z distributions (integrated over all isotopes)
is enhanced for neutron-poor systems \cite{Yang,dago}.
In turn, the staggering on N-distributions (integrated over all 
isotones) is enhanced for neutron-rich systems \cite{lombardo}.
There are also indications for a weakening of the odd-even Z staggering 
with increasing centrality of the collision \cite{dago}.

The odd-even staggering is believed to be a
signature of nuclear structure effects \cite{Ricciardi,dago}.
They may manifest themselves just in the reaction mechanism,
if part of the reaction proceeds through 
low excitation energies \cite{Ademard}.
However, in collisions at intermediate energies the preferred
interpretation is that structure effects are restored in the
final products of hot decaying nuclei and that
the odd-even staggering depends 
- in a complex and presently not very well understood way - 
on the structure of the nuclei produced near the end of the 
evaporation chain \cite{Ricciardi,dago}.

While the odd-even staggering is an interesting topic in itself,
renewed interest has been stirred by recent observations of such phenomena 
in fragments produced in heavy ion collisions at intermediate energies
(15$\alt$E/A$\alt$50 MeV/nucleon).
Indeed, in order to study the symmetry 
energy \cite{colonna,su,raduta} one needs to reliably estimate the 
primary isotopic distributions and this is possible only if the 
effects of secondary decays are small or sufficiently well understood.

This paper 
presents an experimental investigation of the odd-even
staggering in the fragment charge distribution
of the asymmetric collision 
$^{112}$Sn + $^{58}$Ni at 35 MeV/nucleon.
The data extend
the available systematics for the first time
to a higher charge range, from Z=6 to Z$\sim$45.
A self-supporting $^{58}$Ni target 
(200 $\mu$g/cm$^2$ thick) was bombarded by a pulsed $^{112}$Sn beam 
(of $\approx$ 1 ns time resolution),
delivered by the Superconducting Cyclotron of the  
Laboratori Nazionali del Sud of INFN in Catania.
Reaction products were detected
with the \textsc{Chimera} multidetector \cite{Porto,Pagano} 
in the complete configuration consisting of 1192 telescopes.
They are mounted at polar angles from 1$^\circ$ to 176$^\circ$,
in an axially symmetric configuration around the beam axis
with a nominal geometric coverage of $\approx$94\% of $4\pi$.
Each telescope, made of a 300 $\mu$m-thick silicon detector 
followed by a CsI(Tl) scintillator, measures both the deposited 
energy and the time-of-flight of the reaction products
(with flight paths from 3 to 1 m below 30$^\circ$ 
and of 40 cm for larger angles) \cite{Porto,Pagano}.

If the products have enough energy to reach the CsI scintillator, 
they are identified in charge either with the $\Delta$E-E technique, 
or (if they are fast light particles) with the usual fast-slow 
analysis of the CsI signal shape. 
In this way all charges can be individually identified from Z=1 
up to the charge of the projectile (Z=50), with a typical FWHM 
resolution of about 0.4 (0.5) charge units at Z=20 (50),
while isotopic identification is obtained up to Carbon.
The thresholds for Z identification vary
from about 6 MeV/nucleon for Light Charged Particles (LCP, Z$\leq$2) 
to about 23 (27) MeV/nucleon for Ni (Sn) ions.
Below these thresholds, the reaction products are stopped in the 
silicon detectors and for an {\em average} estimate of the charge 
one must rely on the mass, obtained from the energy vs. time-of-flight 
(ToF) correlation.
However, the resolution is limited by the overall ToF resolution 
(beam + detector) and by the length of the available flight-paths.
Moreover, heavy fragments with very low energy need 
important corrections to take into account 
energy losses in dead layers and Pulse Height Defect phenomena.
Finally, the Evaporation Attractor Line (EAL) formula \cite{EAL} is 
used for transforming the mass into an approximate estimate 
of the {\em average} charge of stopped particles
(with a FWHM resolution which for Ni-like quasi-elastic recoils 
is about 10--12 charge units, with non-Gaussian tails).
We anticipate here that particles stopped in the silicon detectors 
are used only in Fig.~2 and not for studying staggering phenomena.
On the contrary,
all charge distributions and the derived information about odd-even
staggering are obtained using only
reaction products which have been ``well-identified-in-charge''
by means of the Si-CsI correlations.
The analysis presented in this paper focuses on events 
where three or more charged products have been detected 
over the whole solid angle. 
This condition removes elastic events and strongly suppresses 
quasi-elastic ones.
At least two of these products are required to be 
fragments with charge Z$\geq$3.

\begin{figure}[tb]
\begin{center}
\includegraphics[width=8.5cm]{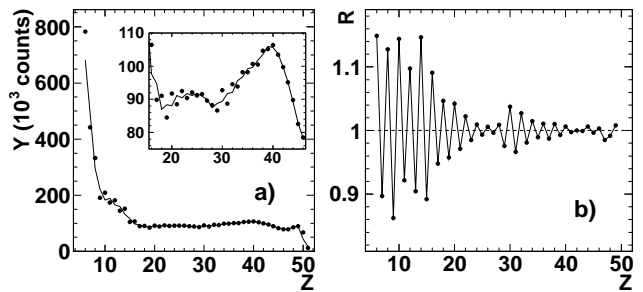}
\caption{(a) Experimental Z distributions (dots) of fragments
  from all measured events.
  The continuous line joins the values $\mathcal{Y}$(Z)
  obtained from a smoothing procedure (see text).
  The inset is an expansion of the region of heavy fragments.
  (b) Ratio R between experimental yields Y(Z) and smoothed 
  values $\mathcal{Y}$(Z) as a function of the fragment charge.
  Statistical errors are smaller than the symbol sizes.}
\label{stagtot}
\end{center}
\end{figure}

The experimental charge distribution of fragments
is shown in Fig. \ref{stagtot}(a).
Aside from a steep rise for light products with Z$\alt$10,
the distribution almost flatly extends up to the projectile charge.
This rather unstructured shape is due the superposition of all 
processes which may occur in the collision.
Since odd-even effects are a rather pervasive feature, 
observed in many different reaction mechanisms, 
some information can be gained already from this inclusive analysis.

A close look at the charge distribution (see inset), shows that some
staggering effect is indeed present, but the large variation of yield,
especially in the low-Z region, hinders a clear appreciation
of its magnitude. 
The smoothing procedure suggested in \cite{dago} was applied, 
which is able to highlight the staggering effect.
For each measured value of the yield Y(Z),
a smoothed value $\mathcal{Y}$(Z) is estimated
with a parabolic fit to the measured yields over five 
consecutive points
(namely, the yields of that Z, and of the two preceding and the 
two following Z's).
The continuous line in Fig. \ref{stagtot}(a)
joins the so obtained values of $\mathcal{Y}$(Z).
The ratio R(Z) between Y(Z) and the smoothed value $\mathcal{Y}$(Z) 
is shown in Fig. \ref{stagtot}(b).
R(Z) oscillates around 1 (dashed line) and
a clear odd-even staggering is observed,
with the even-Z ions being more abundantly
populated than the odd ones.
This 
bears out the gross features already observed in 
previous experiments
\cite{Yang,Winchester,Ricciardi,dago,lombardo}.
The staggering is large (R$\approx$10--15\%) for medium-charge 
ions with Z$\alt$16 and it decreases with increasing Z.
Indeed the rapid disappearance of staggering effects
giving rise to a smooth behavior with increasing mass is a 
common observation, which has been tentatively explained 
with a decrease of the pairing gap and an increasing competitiveness 
of gamma emission with respect to particle decay \cite{Ricciardi}.
It is worth noting that 
previous investigations of staggering effects 
usually stopped around Z=16--20 and none has
covered up to now this large range from Z=6 to Z$\sim$45.
Around Z=30 we observe an enhancement,
with a renewed increase of the yields of the even 
Z=30, 32 fragments with respect to the neighboring odd ones. 
An excess in the production of even elements around Z=30,
similar to that of the present paper,
has been observed also
in the charge distributions of quasi-projectiles measured with 
the \textsc{Fiasco} apparatus \cite{fiasco} for similar systems 
at 30 and 38 MeV/nucleon.
Work on this point is in progress, but 
at present no interpretation of this new ``anomaly'' around Z=30
is at hand.
Trivial effects due to particular features of proton and neutron
separation energies seem unlikely, since the separation energies
do not show any anomalous behavior around this value of Z. 

The inclusive results of Fig. \ref{stagtot} are dominated by
the high yields of peripheral collisions.
For a more detailed analysis, it is necessary to sort the events 
in bins of different centrality and this requires the preliminary 
selection of nearly ``complete'' events.
This was obtained constraining the total reconstructed charge 
Z$_\mathrm{tot}$ and
the total momentum along the beam axis p$_\mathrm{par}$.
For estimating these two variables
--and only for this purpose--
also reaction products stopped in the 
silicon detectors were included.
The mass of non-stopped particles was deduced 
from the measured charge with the EAL formula.

\begin{figure}[tb]
\begin{center}
\includegraphics[width=8.5cm,bb= 30 10 415 240,clip]{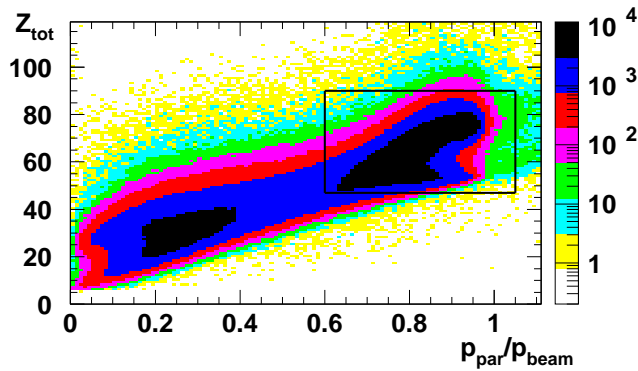}
\caption{(color online)  
  Total detected charge Z$_\mathrm{tot}$ vs. ratio of
  the total longitudinal momentum to the beam momentum
  (particles stopped in the silicon detectors are included).
  The rectangle indicates the ``complete'' events used in the analysis.}  
\label{pztot}
\end{center}
\end{figure}

The correlation between total reconstructed charge and total 
longitudinal momentum is shown in Fig. \ref{pztot}.
The lower branch of the ``fork'' at high p$_\mathrm{par}$
corresponds to events where the Ni-like recoil escaped detection, 
while the two branches at low p$_\mathrm{par}$ correspond to events
lacking detection of the Sn-like in one case, and of both the 
Sn- and Ni-like fragments in the other case. 
The tail of events with Z$>$78 is due to fragments 
(mainly slow Ni-like recoils or target-like fragments) 
stopped in the silicon detector, 
for which no direct charge identification is possible, 
but only an average estimation from their mass.
The adopted acceptance window, indicated by the rectangle in figure,
requires a total reconstructed charge greater than 60\% of the system
charge (Z=78) and a total
longitudinal momentum greater than 60\% of the beam momentum.
Events with low total charge and low total momentum 
were rejected, as they correspond to more incomplete events. 
In this experiment
(one of the firsts with the complete \textsc{Chimera} setup) 
not all telescopes delivered full information or performed equally well. 
Therefore a necessary compromise between exploiting a 
large geometric coverage and having a very good Z resolution 
resulted in using the best 60\% of the detectors (evenly sparse over 
the solid angle) for the analysis of odd-even effects.
The results are insensitive to this choice, since staggering effects 
arise from the comparison of the different production rates of nuclei 
with similar Z values, hitting the same detectors.

\begin{figure}[tb]
\begin{center}
\includegraphics[width=8.2cm,bb= 32 59 575 241, clip]{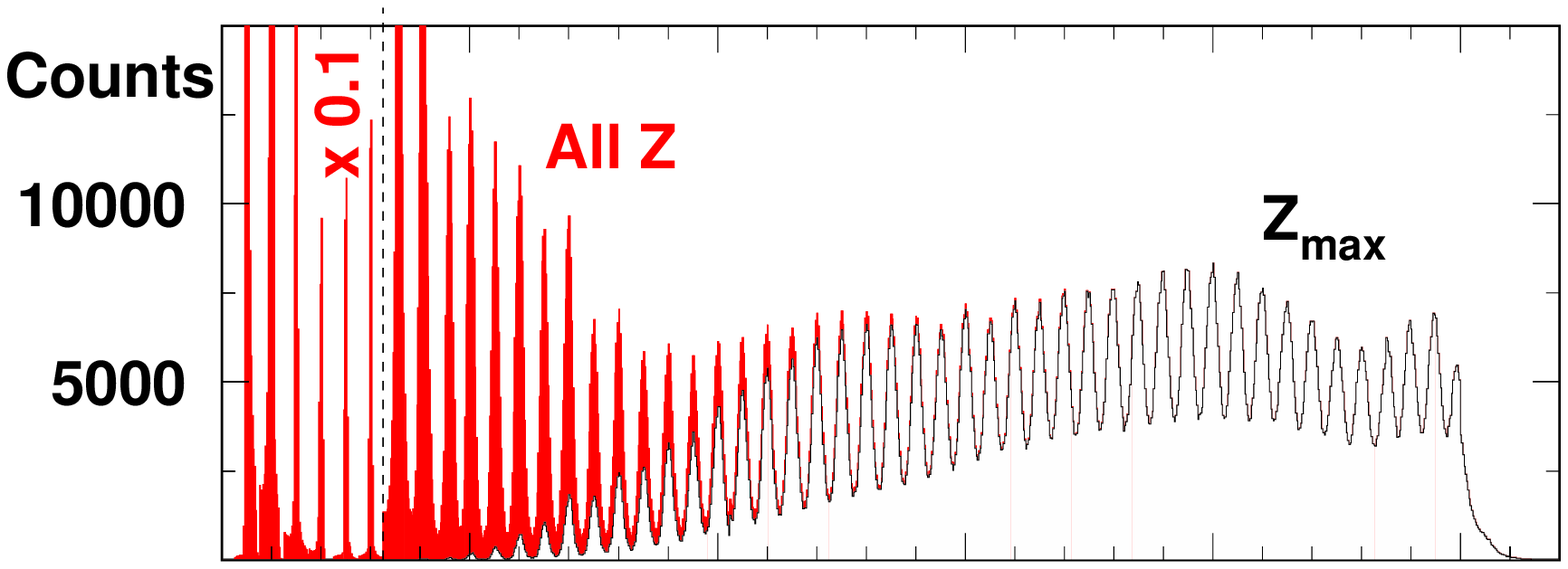}
\includegraphics[width=8.5cm,bb= 5 5 568 346,clip]{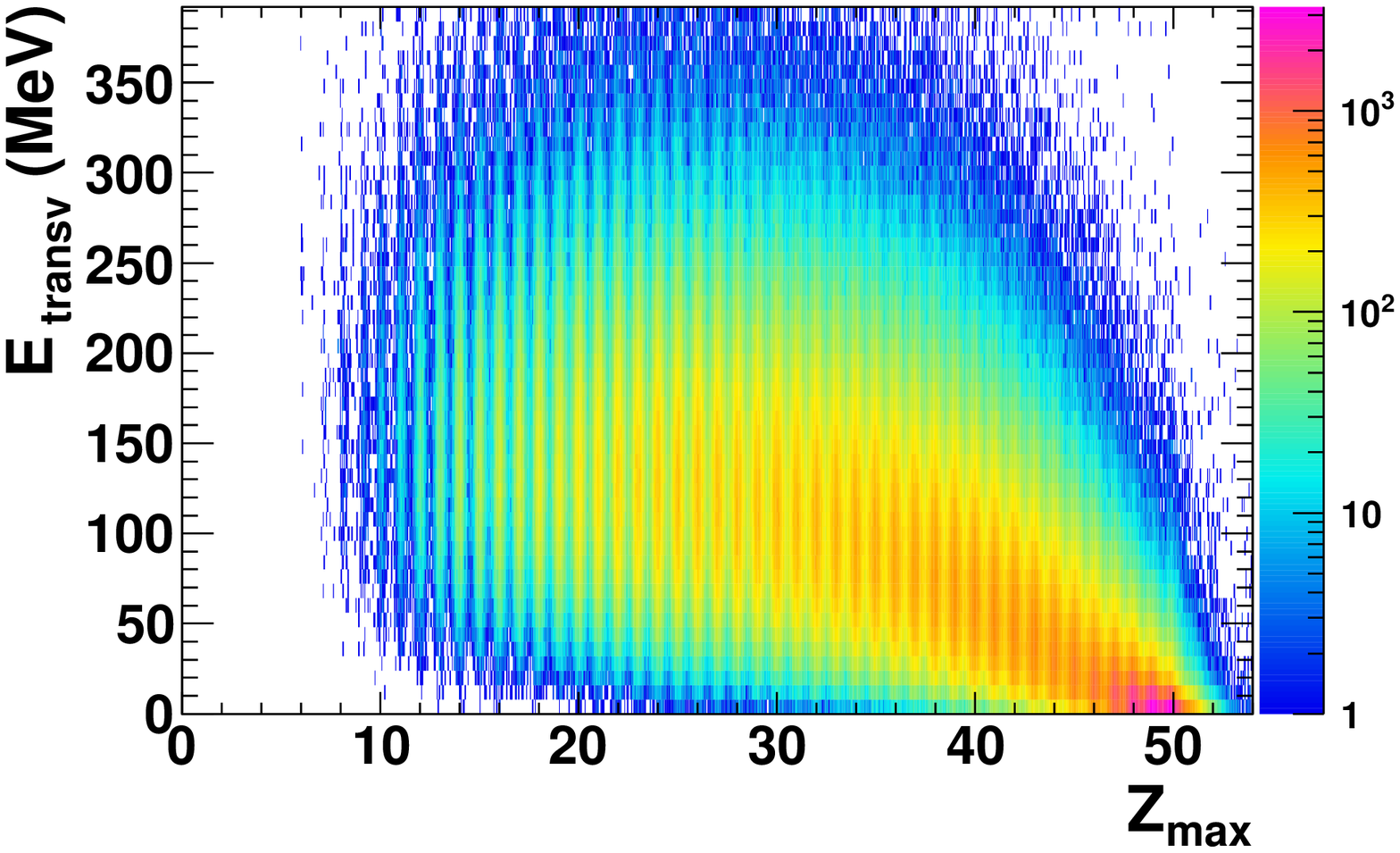}
\caption{(color online)
  Charge distributions for fragments identified in
  charge from the Silicon-CsI correlations. 
  Upper panel: distribution of the heaviest fragment Z$_\mathrm{max}$
  (black histogram) and of all fragments (color histogram),
  with data for Z$\leq$6 downscaled by a factor 10.
  Lower panel: correlation between transverse energy of LCPs
  (E$_\mathrm{transv}$) and Z$_\mathrm{max}$.}
\label{zmaxetrasv}
\end{center}
\end{figure}

The charge distribution and staggering ratio for the selected 
``complete'' events are indistinguishable from those of 
Fig. \ref{stagtot}, apart from the reduction of the statistics 
by a factor of about 2. 
However, it is now possible to find a sorting parameter 
for investigating the evolution of the staggering with the centrality 
of the collision.
In heavy ion collisions at intermediate energies the largest part of the
reaction cross section corresponds to binary events, characterized
by the presence in the exit channel of two heavy remnants, the
quasi-projectile and the quasi-target.
Together with these heavy remnants one observes LCPs and 
Intermediate Mass Fragments 
(IMF, with 3$\leq{\rm Z}\alt$16)
produced by their evaporative decay and possibly by midvelocity emissions 
(see, e.g., \cite{lukasik,fiascoprc74,defilippo}).  
The transverse energy of LCPs was used to estimate the centrality 
of the collisions \cite{lukasik,plagnol}. 
The transverse energy is defined as
E$_\mathrm{transv}$=$\sum_i$ p$_{i \perp}^2$/(2m$_i$), 
where p$_{i \perp}$ 
is the transverse momentum (with respect to the beam axis) 
and m$_i$ the mass of the LCP.
The lower part of Fig. \ref{zmaxetrasv} shows  
the transverse energy E$_\mathrm{transv}$
as a function of the charge of the heaviest fragment Z$_\mathrm{max}$.
In semiperipheral collisions with reverse kinematics the fast,
heavy, forward-going fragment is the quasi-projectile remnant and its
charge is expected to decrease with increasing violence of the 
collision \cite{djerroud,yanez,planeta}.
The good correlation between the two variables
observed in Fig. \ref{zmaxetrasv} indicates that,
although the geometric coverage of the setup was not complete,
the transverse energy of the detected LCPs can be used as an
impact parameter selector.

\begin{figure}[tb]
\begin{center}
\includegraphics[width=8.6cm]{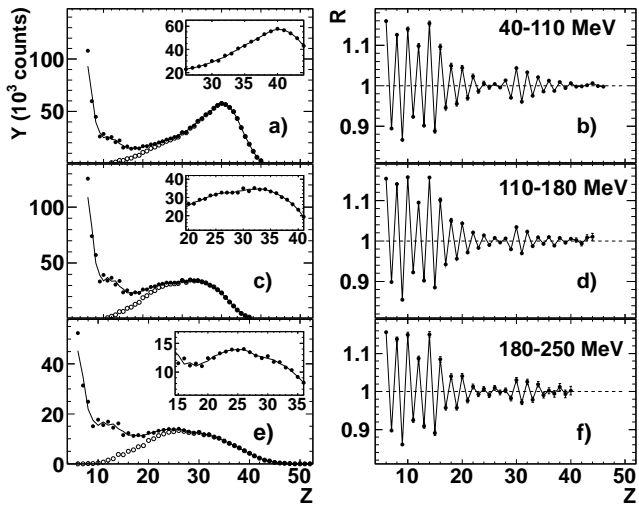}
\caption{Charge distributions of fragments detected in 
  ``complete'' events, for three bins of E$_\mathrm{transv}$: 
  40--110 (a), 110--180 (c) and 180--250 MeV (e). 
  Full points show the yield Y(Z) of all detected reaction 
  products, open points that of the largest fragment
  Y(Z$_\mathrm{max}$). The continuous line joins the values 
  $\mathcal{Y}$(Z) obtained from a smoothing procedure (see text).
  The insets are expansions of the peaks at large Z.
  Statistical errors are smaller than the point size.
  (b), (d), (f) Corresponding ratios R between the experimental 
  yield Y(Z) and the smoothed value $\mathcal{Y}$(Z)
  as a function of Z. 
  Bars represent statistical errors.}
\label{stag}
\end{center}
\end{figure}

The upper part of Fig. \ref{zmaxetrasv} 
shows the projection (black histogram) of the same data 
on the Z$_\mathrm{max}$ axis and, for comparison, also the
Z distribution (color histogram) of all detected particles, 
from protons up to projectile-like fragments.
Although the data are integrated over particle energies and 
summed over all used detectors, one can see a good separation of 
the individual elements up to the highest detected charges.

The experimental charge distributions of fragments
are shown by the full points in Fig. \ref{stag}(a), (c), (e)
for three windows of E$_\mathrm{transv}$.
In each panel one observes a broad peak located at large Z,
which on its left side merges into a very rapidly rising
distribution of light fragments
(the structure of the peak is shown in more detail in the insets).  
With increasing centrality, the position of the peak moves
towards smaller Z values.
This peak in the total distribution is due to the heaviest detected
fragment in the event, as it is clearly shown by plotting the 
distribution of Z$_\mathrm{max}$ (open points).
It is therefore reasonable to ascribe the peak to the quasi-projectile
remnants, which become lighter with increasing centrality
of the collision. 

The staggering ratio R, for the same bins of centrality,  
is shown in Fig. \ref{stag}(b), (d), (f), 
where each error bar indicates the combined statistical error of the
experimental yield and of the smoothed one, fitted to five data points.
Systematic errors are estimated to be two or three times larger
than statistical ones.
The ratio R
appears to be very similar in all three
windows of E$_\mathrm{transv}$.
It is large for IMFs, then it progressively decreases for heavier 
fragments above Z$\sim$16,
with a presently unexplained enhancement around Z=30.
The staggering of IMFs is not only large in size,
but it seems to remain unaltered with increasing centrality.
To better test this last result, we performed
a cleaner selection of central collisions. 
Central collisions, which
represent a small part of the total reaction
cross section, are characterized by a large number of IMFs
and LCPs emitted in events with a nearly spherical shape
(multifragmentation events). 
Therefore they have a rather isotropic distribution of 
flow angle $\theta_\mathrm{flow}$, in contrast to semiperipheral events
that have a distribution peaked around $0^{\circ}$.
Thus one of the usual methods to select 
a clean sample of central-collision events
is to require a large value of $\theta_\mathrm{flow}$
(\cite{indrasilvia} and references therein).
In the present paper we used the selection 
~$\cos(\theta_\mathrm{flow}$)$<$0.6, which corresponds to 
$\theta_\mathrm{flow}>53^\circ$.  
The trends found for central events
resemble those described for semiperipheral collisions.  
The odd-even effect, with the prevalence of even-Z ions, 
remains also for central collisions. 
However, due to the limited cross section and to the reduced fragment 
size in many-fragment events, the observation of the
staggering cannot be extended to as large Z values as for
semiperipheral events.

\begin{figure}[tb]
\begin{center}
\includegraphics[width=8.4cm]{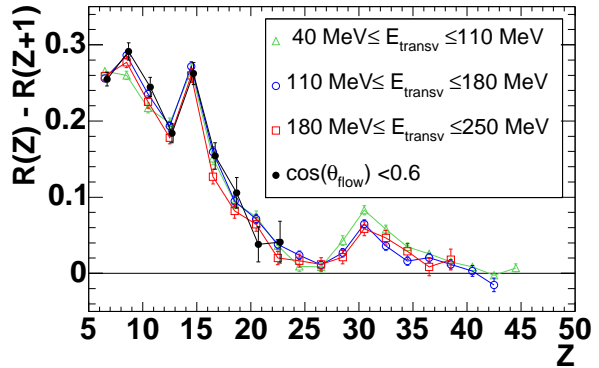}
\caption{(color online) Difference between the staggering ratio for 
  {\it even} Z values and for the following {\it odd} Z+1 values, 
  R(Z)-R(Z+1), for different centrality selections based on conditions
  on ~$\cos(\vartheta_\mathrm{flow}$) or E$_\mathrm{transv}$ (see text).  
  Symbols, with statistical errors, are drawn at the 
  semi-integer Z of each even-odd pair.}
\label{cfr}
\end{center}
\end{figure}

The possible dependence of the odd-even staggering on
the centrality of the reaction is summarized in Fig. \ref{cfr}.
Here each point, 
drawn at the semi-integer value Z+$\frac{1}{2}$ 
of an even-odd pair,
represents the difference
between the staggering ratio R(Z) for that even Z and the ratio R(Z+1)
for the following odd Z+1 value.
The different symbols refer to the different selections 
of E$_\mathrm{transv}$ in Fig. \ref{stag} and to the central
collisions.
From Fig. \ref{cfr} three points are worth noting.
First, all distributions are very well superimposed on each other,
with practically no dependence on impact parameter
(at least within the sensitivity of our measurement).
This seems
at variance with the weak dependence on centrality 
found in a lighter system \cite{dago}.
Second, the charge distributions of IMF (3$\leq{\rm Z}\alt$16) 
display a strong even-odd effect, a feature which is present in all
previous experiments which have investigated staggering phenomena.
Also the large opposite behavior of the yields of 
Z=8 with respect to Z=9 (and of Z=14 with respect to Z=15) 
is visible in previous data \cite{Ricciardi}.
Finally, the staggering effect decreases rapidly with increasing 
charge of the heavier fragments, except for an unexpected moderate
enhancement around Z=30, which was not observed previously
because of the limited range of investigated charges.

Past attempts to explain staggering effects on the base of nuclear
structure effects could reproduce the experimental data only in a 
qualitative, but not in a quantitative way.
This led to suppose that not only the available phase space at the 
end of the evaporation process is important, but also the number 
of available levels in the mother nucleus may play a role \cite{Ricciardi}.
A more recent investigation assumes that it may be necessary to have 
a good knowledge not only of the last decay step, but of the last 
few ones \cite{dago2}, particularly for what concerns 
``pairing and isospin effects in the level density'' 
at rather high excitation energies.
  
The selections adopted in the present paper
correspond to events with different 
impact parameters (from peripheral to central), therefore the
observation that the staggering amplitude of IMFs is independent of 
the reaction class demonstrates that this phenomenon 
keeps almost no memory of
the preceding reaction dynamics.
In fact, IMFs may be produced by different mechanisms,
ranging from the possible multifragmentation of a ``central'' source to 
the decay of a hot quasi-projectile, via evaporation or sequential fission.
This in turn supports the idea that the odd-even staggering of IMFs
arises in the last few steps of their decay and may be linked
to details of 
the structure of the involved nuclei \cite{Ricciardi,dago2}.

In summary, we have presented experimental data from the 
reaction $^{112}$Sn+$^{58}$Ni at 35 MeV/nucleon, collected with the 
\textsc{Chimera} multidetector during one of its first campaigns in
complete configuration. 
The presented results focus on the charge distributions of fragments 
which are well identified in charge Z.
We have shown, also thanks to a smoothing technique,
that it is possible to
evidence a clear odd-even staggering effect in the charge distribution 
for light-to-medium fragments, up to values around Z$\sim$30:
even-Z fragments are produced more abundantly than odd-Z ones.
The amplitude of the
staggering effect decreases with increasing Z.
There is an ``anomaly'' in the region around Z=30--32, where the
difference in the yields between even and odd Z nuclei increases again.
This fact is as yet not understood and deserves further investigations. 

The experimental data have been sorted in samples of increasing
centrality, from peripheral to central collisions.
From our analysis the odd-even effects persist
independently of the centrality of the reaction.
This suggests that staggering is little sensitive to the
preceding dynamics and has to be ascribed mainly to the last steps in
the decay chain of hot fragments.
The current tentative interpretation is that they may be related 
to details of the internal structure of the near-final fragments
\cite{Ricciardi,dago}.
From an experimental point of view, the investigation of
these phenomena will benefit 
from the availability of new radioactive beams and
from the next generation of detectors \cite{carboni},
capable of extending mass and charge 
identification (with unit resolution)
to medium-high Z nuclei, over large solid angles.
This will allow to access more exotic systems and to 
investigate staggering effects in detail,
as a function of both Z and N.

\begin{acknowledgments}
The authors are indebted to the staff of the Superconducting Cyclotron
at LNS, in particular D.\ Rifuggiato, for providing
a very high quality beam. 
The authors also wish to warmly thank M. D'Agostino for many stimulating 
discussions and for careful reading of the manuscript.
\end{acknowledgments}

\end{document}